\documentclass[usenatbib]{mnras}

\usepackage{newtxtext,newtxmath}
\usepackage[T1]{fontenc}
\usepackage{ae,aecompl}

\usepackage{mathtools}
\usepackage{siunitx}
\usepackage{amssymb,multirow,tabularx}
\usepackage{amsmath}
\usepackage{graphicx}
\usepackage{comment}
\usepackage[normalem]{ulem}
\usepackage{BibDef}
\usepackage{xcolor}
\usepackage{hyperref}
\hypersetup{
	colorlinks=true,        % false: boxed links; true: coloured links
	linkcolor=blue,         % colour of internal links
	citecolor=blue,         % colour of links to bibliography
}
\newcommand{\msun}{~\rm M_{\large \odot}}
\newcommand{\ud}{~{\rm d}}

\title[DF in rotating discs]{Dynamical friction-driven orbital circularisation in rotating discs: a semi-analytical description}

\author[M. Bonetti et al.]{Matteo Bonetti,$^{1,2,3}$ Elisa Bortolas,$^4$ Alessandro Lupi,$^5$ Massimo Dotti,$^{1,2,3}$
\newauthor and Sandra I. Raimundo$^{6}$\\
$^1$Dipartimento di Fisica ``G. Occhialini'', Universit\`a degli Studi di Milano-Bicocca, Piazza della Scienza 3, I-20126 Milano, Italy\\
$^2$INFN, Sezione di Milano-Bicocca, Piazza della Scienza 3, I-20126 Milano, Italy\\
$^3$INAF, Osservatorio Astronomico di Brera, Via E. Bianchi 46, I-23807, Merate, Italy\\
$^4$Center for Theoretical Astrophysics and Cosmology, Institute for Computational Science, University of Z\"urich, Winterthurerstrasse\\ \ 190, CH-8057 Z\"urich, Switzerland\\
$^5$Scuola Normale Superiore, Piazza dei Cavalieri 7, I-56126 Pisa, Italy\\
$^6$DARK, Niels Bohr Institute, University of Copenhagen, Lyngbyvej 2, 2100 Copenhagen, Denmark\\
}

\date{Accepted 2020 April 3. Received 2020 March 30; in original form 2020 February 4}

\pubyear{2020}

\begin{document}
\label{firstpage}
\pagerange{\pageref{firstpage}--\pageref{lastpage}}
\maketitle

%%%%%%%%%%%%%%%%%%%%%
\begin{abstract}
  We present and validate a novel semi-analytical approach to study the effect of dynamical friction on the orbits of massive perturbers in rotating stellar discs.
  We find that dynamical friction efficiently circularises the orbit of co-rotating perturbers, while it constantly increases the eccentricity of counter-rotating ones until their angular momenta reverse, then once again promoting circularisation.
  Such ``drag toward circular corotation'' could shape the distribution of orientations of kinematically decoupled cores in disc galaxies, naturally leading to the observed larger fraction of co-rotating cores.
\end{abstract}
%%%%%%%%%%%%%%%%%%%%%


%%%%%%%%%%%%%%%%%%%%%%%%%%%%%%%%%%%%%%%%%%%%%%%%%%%%%%%%%%%%%%%%%%%%%%%%%%%%%%%%%%%%%
%%%%%%%%%%%%%%%%% BODY OF PAPER %%%%%%%%%%%%%%%%%%
\section{Introduction}
\label{sec:intro}

The orbital evolution of massive perturbers (MPs, i.e. objects
considerably heavier than the single bodies forming the galactic
structure within which the perturber is moving) under the effect of
dynamical friction \citep[DF, see][]{BT} has been studied in
many different contexts and over a broad range of spacial scales: from
the pairing and mergers of galaxies at kpc scales \citep[see][for a recent review]{derosa20}, through the sinking and tidal disruption
of globular clusters in the central kpc of galaxies \citep{BT}, down to the
formation of massive black hole (MBH) binaries at pc-scales \citep[MBHBs, see e.g.][for a review]{DSD12}.

The interaction between a MP and a rotating system embedding it
(either stellar or gaseous) has been found to reverse the direction of
the MP angular momentum if it is initially anti-aligned with that of
the background, or, if the two are aligned, to circularise the MP
orbit. Such behaviour has first been observed by \cite{Dotti06} in
simulations of MBHs in circumnuclear gaseous discs, common in
ultra-luminous infrared galaxies \citep{SM96, DS98}, and observed to
form in gas rich galaxy mergers \citep{Barnes02, Mayer07, CD17}. This
behaviour, dubbed here ``drag toward circular corotation''\footnote{\label{footnote:1} The angular momentum reversal of initially retrograde orbits was originally referred to as ``angular momentum flip'' \citep[e.g.][]{Dotti06}. This convention could be misleading, since the orbital plane of the orbit does not evolve gradually passing through a perpendicular configuration. For this reason we decided to adopt the new conventional name.} (DTCC
hereafter), has been demonstrated to be independent of the nature of
the background, working in nuclear stellar discs as well
\citep{Dotti07}, and then observed to be present in larger discs during
galaxy minor mergers \citep{Callegari11, Fiacconi13}. \cite{Dotti06}
proposed a qualitative model in which DF is
responsible for the DTCC, by accelerating the MP in the direction
opposite to its instantaneous motion relative to the local
environment. DF always decelerates counter-rotating MPs, increasing
their eccentricity without changing the orbital plane, until the DF
effect integrated over one orbit suffices to reverse the MP motion; DF
instead increases the angular momentum of co-rotating MPs at the
apocentre (where the MP is slower than the background) and decreases
it at the pericentre, circularising the orbit.

Contrary to the cases of DF exerted by spherical structures with
isotropic velocity fields \citep[e.g.][]{Chandra43}, an analytical
description of DF in composite systems with at least one rotating
disc-like structure (such as a disc galaxy, or a nuclear disc embedded
in the larger bulge) has not been developed yet. As a consequence, the
model put forward by \cite{Dotti06} and adopted as physical
explanation for the DTCC by the all the above-mentioned studies has not been
proved to date. Such proof is long due, because of the numerous and
important consequences of the DTCC, among
which: the slow-down of the MP orbital shrinking \citep[see e.g. the
discussion in][]{Mayer13}, the enhanced accretion onto orbiting MBHs
once they circularise \citep{Dotti09, Callegari11}, the
alignment of the spins of MBHs in circumnuclear discs, well before
they bind in a binary \citep{Dotti10} and the consequent low
recoils expected at coalescence \citep[but for a small tail of fast
recoiling remnants, see][]{Lousto12}.

In this paper we describe a simple analytical form for the dynamical friction on MPs in rotating systems (Section~\ref{sec:method}), and integrate it numerically in composite systems including discs. We compare the outcome of this new semi-analytical description with the results of high resolution N-body simulations of the same galactic models (Section~\ref{sec:results}). Finally, as discussed in Section~\ref{sec:discussion}, the agreement between the two numerical tests demonstrates that DF is indeed the only responsible for the DTCC in rotating systems. In the same section we comment on the possible implications of the DTCC for kinematically decoupled cores, and discuss some future expansions and applications of our model.

%%%%%%%%%%%%%%%%%%%%%%%%%%%%%%%%%%%%%%%%%%%%%%%%%%%%%%%%%%%%%%%%%%%%%%%%%%%%%%%%%%%%%
\section{Method}
\label{sec:method}

Employing a C++ implementation of the Bulirsch-Stoer \citep{Bulirsch1966} integration scheme,\footnote{See e.g. \citet{Bonetti2016} for further details about the integration algorithm.} we numerically integrate the differential equations describing the motion of a MP when subjected to the global conservative gravitational potential $\phi$ generated by the galaxy mass distribution, as well as because of the action of DF.

In order to obtain analytical equations of motion, we model the host galaxy through appropriate density-potential pairs ($\rho$/$\phi$), in particular we choose:

%%%%%%%%%%%%%%%
\begin{itemize}
    \item a Navarro-Frenk-White profile \citep[NFW,][]{Navarro1997}, with mass $M_h$ and scale radius $a_h$, to describe the dark matter (DM) halo
    %%%%%%%%%%%%%%
    \begin{align}
        \rho_h(r) &= \dfrac{M_h}{4 \pi a_h^3} \dfrac{a_h}{r (1+r/a_h)^2},\\
        \phi_h(r) &= -\dfrac{G M_h}{r}\ln\left(1+\dfrac{r}{a_h}\right);
    \end{align}
    %%%%%%%%%%%%%%

    \item an Hernquist profile \citep{Hernquist1990}, with mass $M_b$ and scale radius $a_b$, to model a compact stellar bulge 
    %%%%%%%%%%%%%%%%
    \begin{align}
        \rho_b(r) &= \dfrac{M_b}{2\pi} \dfrac{a_b}{r (r+a_b)^3},\\
        \phi_b(r) &= -\dfrac{G M_b}{r+a_b};
    \end{align}
    %%%%%%%%%%%%%%

    \item an exponential disc profile \citep[see e.g.][]{BT}, with mass $M_d$, scale radius $R_d$ and scale height $z_d$, to represent the galactic stellar disc
    %%%%%%%%%%%%%%%%
    \begin{align}
        \rho_d(R,z) &= \dfrac{M_d}{4\pi R_d^2 z_d} {\rm e}^{-R/R_d} {\rm sech}^2\left(\dfrac{z}{z_d}\right),\\
        \phi_d(R,0) &\approx -\dfrac{G M_d}{2 R_d^2} R \left[I_0(w) K_1(w) -I_1(w) K_0(w) \right]\nonumber\\
        &+ \dfrac{G M_d z_d}{R_d^2}{\rm e}^{-R/R_d},\label{eq:pot_disk}
    \end{align}
    where $R = \sqrt{x^2+y^2}$, $w = R/(2 R_d)$, while $I,K$ are modified Bessel functions of the first and second kind respectively. The disc potential is known only in the disc equatorial plane, where the orbits of the perturbers are set. We note that the first term in equation~(\ref{eq:pot_disk}) is strictly valid for a razor-thin disc, while the second term is an approximate correction that takes into account the non-zero thickness of the exponential disc \citep[see e.g.][]{Kuijken1989}.
    %%%%%%%%%%%%%%
\end{itemize}
%%%%%%%%%%%%%

Under the above assumptions, the total mass density and total gravitational potential (in the $z=0$ plane) are simply given by
%%%%%%%%%%%%%%%%
\begin{align}
    \rho(\mathbf{r}) &= \rho_h(r)+\rho_b(r)+\rho_d(R,z),\\
    \phi(\mathbf{r}) &= \phi_h(r)+\phi_b(r)+\phi_d(R,0),
\end{align}
%%%%%%%%%%%%%%%%
and the particle $m_p$ at position $\mathbf{r}_p = (x_p,y_p,z_p)$ with velocity $\mathbf{v}_p = (v_{p_x},v_{p_y},v_{p_z})$ is therefore subjected to a conservative acceleration 

%%%%%%%%%%%%%%%%
\begin{equation}
    \mathbf{a}_{\rm cons} = -\nabla \phi.
\end{equation}
%%%%%%%%%%%%%%%%

In addition to the conservative gravitational force, the motion of the MP is also affected by the ``dissipative'' DF force.\footnote{DF is not dissipative in nature, but here we do not consider the effect that the perturber exerts on the surrounding matter, changing its density and velocity distributions, and focus only on the dynamical evolution of the perturber itself.}
In the case of spherically symmetric non-dissipative systems, the effect of DF is often modeled via the \citet{Chandra43} formula:

%%%%%%%%%%%%%%%%%%
\begin{equation}
    \label{eq:DF_sph}
    \mathbf{a}_{\rm df,sph} = -2\pi G^2 \rho(r) \ln(1+\Lambda^2) m_p \left({\rm erf}(X) - \dfrac{2 X {\rm e}^{-X^2}}{\sqrt{\pi}}\right)  \dfrac{\mathbf{v}_p}{v_p^3}
\end{equation}
%%%%%%%%%%%%%%%%%%
where $X = v_p/(\sqrt{2} \sigma)$ is a dimensionless parameter relating the perturber velocity to the isotropic velocity dispersion $\sigma$, while  $\ln(1+\Lambda^2)$ (or more frequently $\ln \Lambda$) is commonly known as the Coulomb logarithm, with $\Lambda = p_{\rm max}/p_{\rm min}$ being the ratio between the maximum and minimum impact parameters.
The above expression for DF is derived in the simplistic assumption of a MP moving in an infinite and homogeneous density background whose velocity distribution is strictly Maxwellian. 
In spite of this, equation~(\ref{eq:DF_sph}) has been proven to reproduce the inspiral of a MP surprisingly well for a vast range of spherically symmetric and isotropic density profiles, once the radius-dependent local velocity dispersion $\sigma(r)$ is plugged in the equation \citep[see, e.g.][]{Tremaine1984}.

Recently, a series of investigations \citep{Hashimoto2003, Just2005, Just2011, Petts2015, Petts2016}  showed that the adoption of position dependent minimum and maximum impact parameters further improves the agreement with MP orbital decays obtained in full $N$-body simulations. 
Motivated by this, we implement equation~(\ref{eq:DF_sph}) separately for the halo and bulge components, adopting position-dependent velocity dispersion $\sigma(r)_{x}$ and mass density $\rho(r)_x$ (where $x$ refers either to the halo or to the bulge) and the following expressions for the maximum and minimum impact parameters: 

%%%%%%%%%%%%%%%%%%
\begin{align}\label{eq:finesse}
    p_{\rm max, x} &= r/\gamma_x, \nonumber \\
    p_{\rm min} &= \max(G m_p v_p^{-2}, D_p), \nonumber \\ 
    \gamma_x &= - \dfrac{\ud \ln \rho_x}{\ud \ln r}
\end{align}
%%%%%%%%%%%%%%%%%%
where $\gamma_x$  represents the logarithmic slope of the density profiles of the two spherical components, $r$ is the radial coordinate and $D_p$ is the physical radius of the MP, which is zero for the case of an MBH.  
Note that, in the case of multiple component systems, as the one we present in this paper, the value of $\sigma(r)_x$ should be formally derived self-consistently accounting for all components. 
However, the comparison with the results of a full N-body simulation discussed in the next section demonstrates that, for the parameters we consider, our simple approximation is sufficient to recover the main features of the perturber orbital evolution.

Contrary to the isotropic cases, the DF acceleration due to a rotating disc component is not necessarily opposed to the perturber direction of motion with respect to the centre of mass of the whole system, but, as discussed in the introduction, primarily depends on the direction of the relative velocity vector between the perturber and the surrounding medium. In fact, as long as the orbit is counter-rotating or, if co-rotating and eccentric, DF acts instantaneously in the direction opposite to the velocity of the perturber with respect to the surrounding matter in its proximity, making the dependence on the velocity direction more relevant than that on the specific shape of the density profile.\footnote{Note that for circular co-rotating orbits in thin discs (i.e. when the support due to any pressure gradient is negligible), the perturber would be at rest with the matter in its immediate proximity, and any orbital evolution would be driven by the opposite forces exerted by matter inside and outside the perturber orbit \citep[see e.g. the discussion in][]{Mayer13}}
Being interested in confirming the DTCC effect, we limit our investigations to non-circular co-rotating cases, and therefore we make the simplifying assumption that the DF in the disc is primarily determined by the local distribution of matter and velocity around the MP. In particular, in the expression:

%%%%%%%%%%%%%%%%
\begin{equation}
    \dfrac{\ud \mathbf{v}_p}{\ud t} = -2\pi G^2 m_p \ln(1+\Lambda^2) m_a \int \ud^3 \mathbf{v}_a f(\mathbf{v}_a) \dfrac{\mathbf{v}_p-\mathbf{v}_a}{|\mathbf{v}_p-\mathbf{v}_a|^3}
\end{equation}
%%%%%%%%%%%%%%
we assume that the distribution function $f(\mathbf{v}_a)$ is a delta function around the circular velocity at the position of the MP, i.e.

%%%%%%%%%%%%%%%%
\begin{equation}
    f(\mathbf{v}_a) = \delta^3( \mathbf{v}_a-\mathbf{v}_{\rm circ}(R_p) )
\end{equation}
%%%%%%%%%%%%%%%%
yielding

%%%%%%%%%%%%%%%%
\begin{equation}
    \label{eq:DF_disk1}
    \dfrac{\ud \mathbf{v}_p}{\ud t} = -2\pi G^2 m_p \ln(1+\Lambda^2) \rho_d(R_p,0) \dfrac{\mathbf{v}_p-\mathbf{v}_{\rm circ}(R_p)}{|\mathbf{v}_p-\mathbf{v}_{\rm circ}(R_p)|^3},
\end{equation}
%%%%%%%%%%%%%%
where $\rho_d(R_p,0)=m_af(\mathbf{v}_{\rm circ}(R_p))$, while $\mathbf{v}_{\rm circ}$ is given by

%%%%%%%%%%%%%%%%
\begin{equation}
   |\mathbf{v}_{\rm circ}(R_p)|^2 = R \dfrac{\partial \phi}{\partial R}\biggl|_{R=R_p}. 
\end{equation}
%%%%%%%%%%%%%%%%
Due to the paucity of detailed studies on dynamical friction in rotating systems, a set of equations like~\eqref{eq:finesse} for discs has not been discussed in literature so far.

Nevertheless, in order to obtain a simple form for the acceleration, we introduce a physically motivated prescription, based on the Coulomb logarithm, similar in spirit to the spherical case. In particular, we define $\Lambda = p_{\rm max, d}/p_{\rm min, d}$, with the maximum impact parameter taken equal to the scale-height of the disc, as the underlying assumption of quasi-homogeneous three-dimensional density fails on scales larger than $z_d$. The minimum impact parameter is defined instead as an effective influence radius based on the relative velocity $v_{\rm rel} = |\mathbf{v}_p-\mathbf{v}_{\rm circ}|$, i.e.
%%%%%%%%%%%%%%%%%%
\begin{align}\label{eq:finesse_disk}
    p_{\rm min, d} &= \dfrac{G m_p}{v_{\rm rel}^2 + 0.01 v_{\rm circ}^2}.
\end{align}
%%%%%%%%%%%%%%%%%%
The additional term $0.01 v_{\rm circ}^2$ at the denominator is added only for purely operational purposes. It represents a numerical floor to $p_{\rm min,d}$ that avoids its divergence and therefore a spurious suppression of the DF force due to the disc once the circular velocity is met.\footnote{Actually, if $p_{\rm min, d}$ diverges then $\Lambda^2 \rightarrow 0$ making $\ln(1+\Lambda^2)$ vanish. We checked that our setup is practically insensitive to any change up to an order of magnitude, both up and downwards, of the factor suppressing $v_{\rm circ}$.} 

In addition to the above prescription and trying to account for the other possible dependencies of DF that cannot be accounted for in this simple framework (e.g. the faster decrease of density in the vertical direction w.r.t. the radial one), we introduce an additional tunable constant $A_{\rm disc}$ in equation~(\ref{eq:DF_disk1}), yielding to
%%%%%%%%%%%%%%%%%%
\begin{equation}\label{eq:DF_disk22}
    \mathbf{a}_{\rm df,disc} = -A_{\rm disc} 2\pi G^2 \ln(1+\Lambda^2) m_p \rho_d(R,0) \dfrac{\mathbf{v}_p-\mathbf{v}_{\rm circ}(R)}{|\mathbf{v}_p-\mathbf{v}_{\rm circ}(R)|^3}.
\end{equation}
%%%%%%%%%%%%%%%%%%
The value of the constant $A_{\rm disc}$ is determined by comparing the semi-analytical orbital evolution of the perturber with that obtained in an N-body simulation, as discussed in Section~\ref{sec:results}.

Finally, including all terms presented above, the total acceleration acting on $m_p$ reads

%%%%%%%%%%%%%%%%%%
\begin{equation}
    \mathbf{a} = \mathbf{a}_{\rm cons} + \mathbf{a}_{\rm df,sph} + \mathbf{a}_{\rm df,disc},
\end{equation}
%%%%%%%%%%%%%%%%%%
which allows us to numerically integrate the orbital motion in the galactic disc.

%%%%%%%%%%%%%%%%%%%%%%%%%%%%%%%%%%%%%%%%%%%%%%%%%%%%%%%%%%%%%%%%%%%%%%%%%%%%%%%%%%%%%
\section{Results: Comparison with N-body simulations}
\label{sec:results}

%%%%%%%%%%%%%%%%%%
\begin{figure}
    \centering
    \includegraphics[width=\columnwidth]{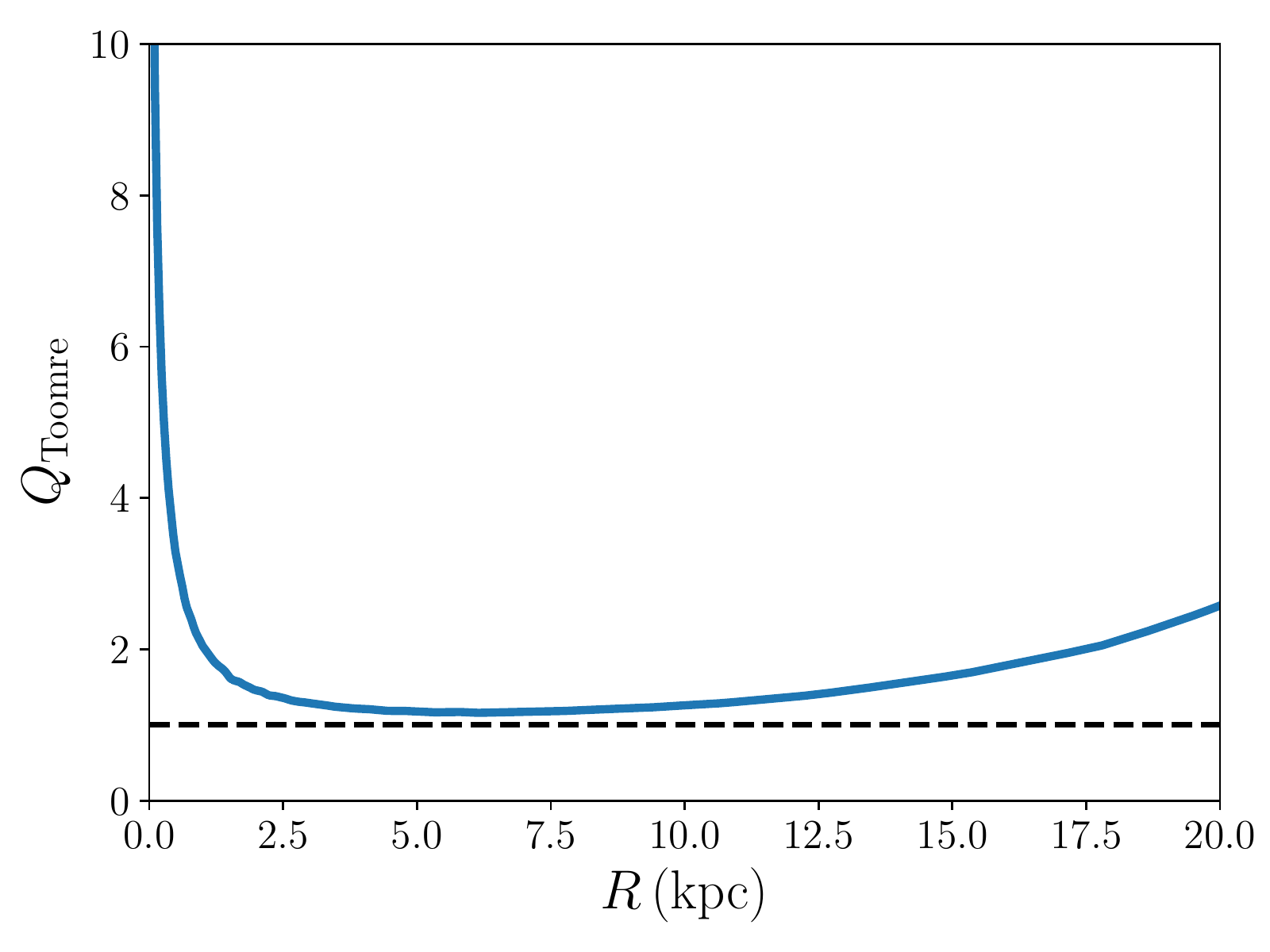}
    \caption{Toomre $Q$ parameter for the stellar disc component in our galaxy model ICs. The black dashed line corresponds to the minimum $Q$ to avoid strong non-axisymmetric perturbations developing in the disc.}
    \label{fig:ToomreQ}
\end{figure}
%%%%%%%%%%%%%%%%%%

%%%%%%%%%%%%%%%%%
\begin{table}
    \centering
    \begin{tabular}{cccc}
        \hline\hline
        & Halo & Bulge & Disc \\
        \hline\\
        Mass & $1.1 \times 10^{12}\msun$ & $2.2\times 10^{9}\msun$ & $4.4\times 10^{10}\msun$ \\
        Scale Radius & $37 \ (21)$~kpc & $0.96$~kpc & $4.25$~kpc \\
        Scale Height & -        & - & $0.85$~kpc \\
        Profile & Hernquist (NFW) & Hernquist & Exponential\\ 
        $N_{\rm part}$ & $10^6$ & $5\times 10^5$ & $2\times 10^6$\\
        $\varepsilon$ & 40~pc & 10~pc & 10~pc\\
        \hline\hline
    \end{tabular}
    \caption{Structural and numerical parameters of the three galaxy components: halo, bulge and disc. For the DM halo, \textsc{GalIC} assumes an Hernquist profile with the scale radius set by the concentration parameter of the halo \citep[assumed $c=10$; see][for full details]{yurin14}. $N_{\rm part}$ and $\varepsilon$ correspond to the number of particles and the Plummer-equivalent gravitational softening employed for each component respectively.}
    \label{tab:paramters}
\end{table}
%%%%%%%%%%%%%%%%%

To validate our analytic prescription we perform high-resolution N-body simulations of an isolated galaxy with the code \textsc{gizmo} \citep{hopkins15}, descendent of \textsc{gadget2} \citep{springel05}, where gravity is solved via a multipole hierarchical method based on a Barnes-Hut Oct-Tree structure, that guarantees an optimal balance between accuracy and performance. 

Our galaxy model (used both in the semi-analytical calculation and in the N-body runs) includes a dark matter halo, described in the semi-analytical framework by a NFW profile and modeled as a   \citet{Hernquist1990} model in the N-body run. In the latter case,  the scale radius is set according to the halo concentration parameter, assumed to be $c=10$. A stellar exponential disc and a central spherical bulge (also modelled as a spherical Hernquist profile) are embedded in the DM component. The model parameters are summarized in Tab.~\ref{tab:paramters}. 

%%%%%%%%%%%%%%%%%%
\begin{figure*}
    \centering
    \includegraphics[width=0.48\textwidth]{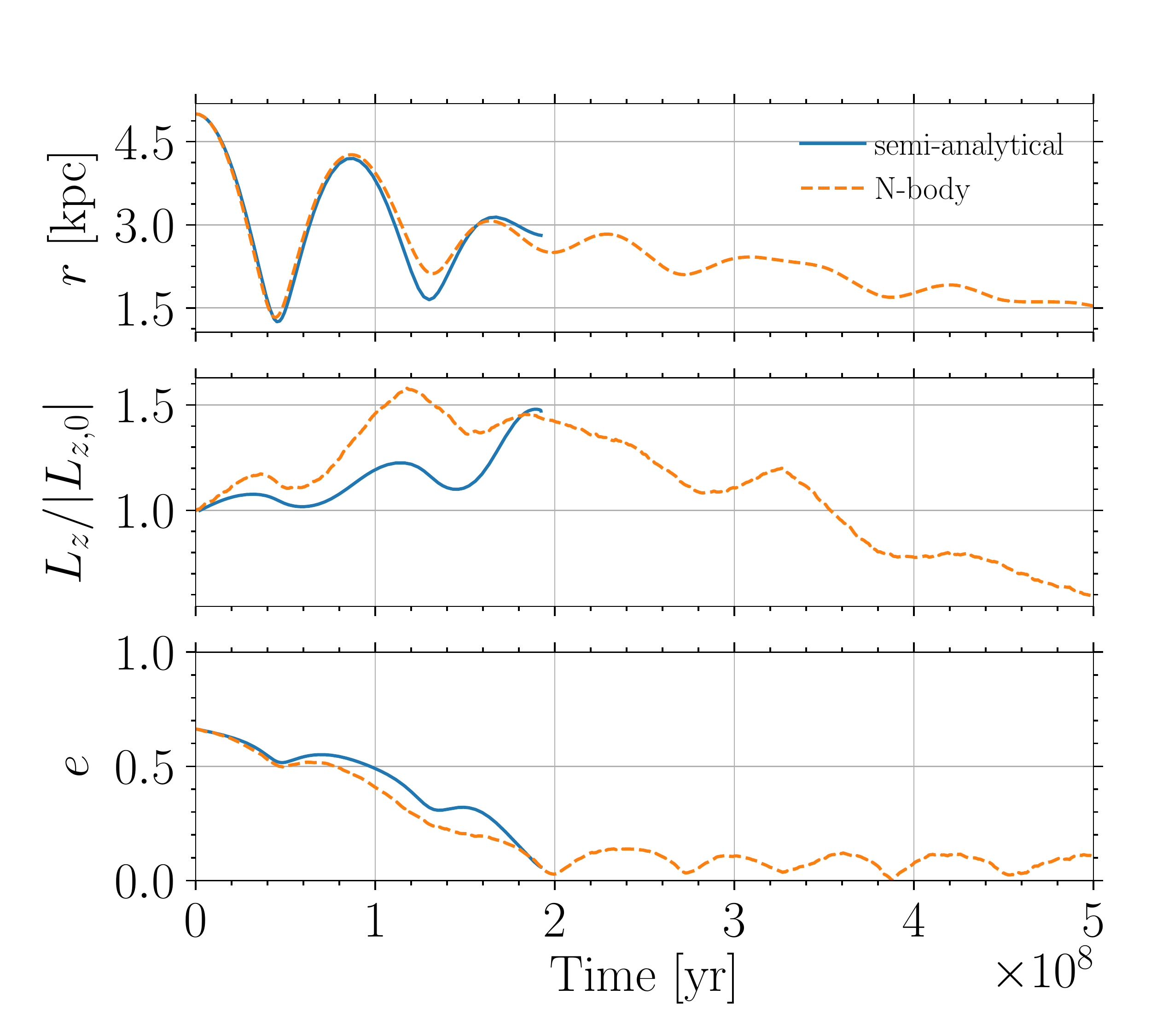}
    \includegraphics[width=0.48\textwidth]{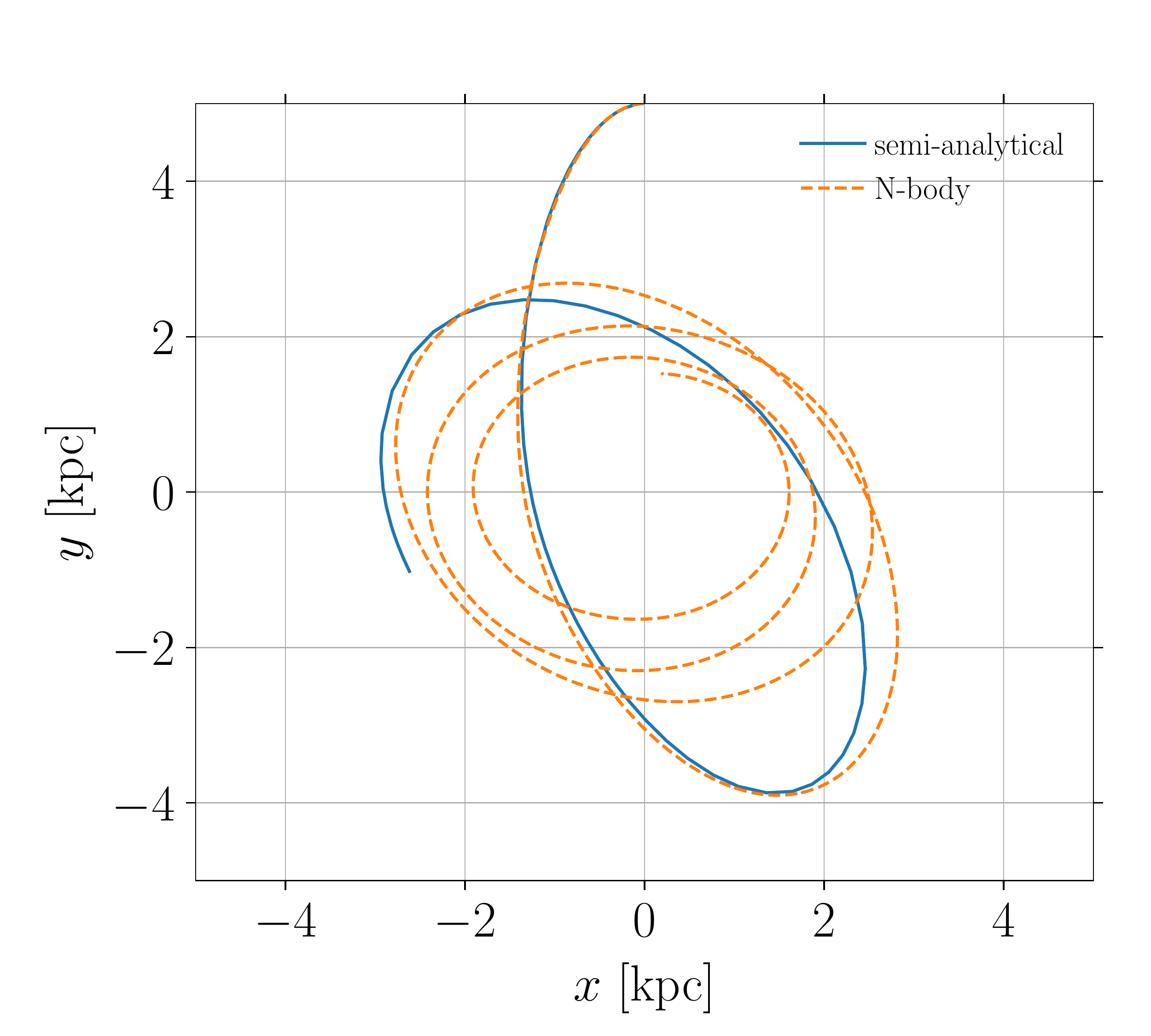}
    \caption{Evolution of an initially prograde perturber. The left-hand panels, from top to bottom, refer respectively to the perturber separation, angular momentum (normalized to its initial magnitude) and eccentricity as a function of time. The right-hand panel shows the perturber orbit in the galactic equatorial plane.
    In all panels, our semi-analytical approach is shown as a blue solid line and the N-body run as an orange dashed line.}
    \label{fig:prograde}
\end{figure*}
%%%%%%%%%%%%%%%%%%
%%%%%%%%%%%%%%%%%%
\begin{figure*}
    \centering
    \includegraphics[width=0.48\textwidth]{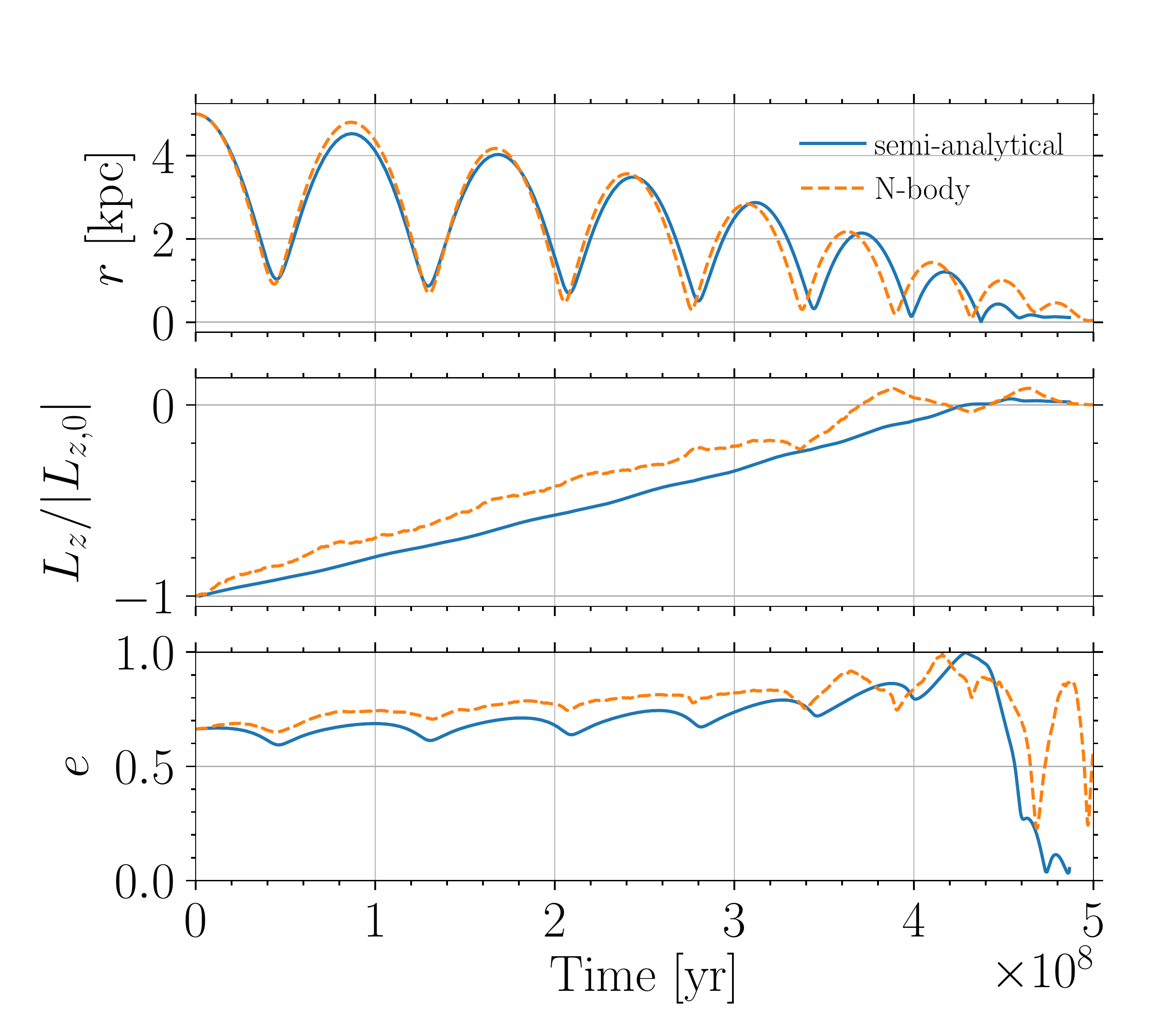}
    \includegraphics[width=0.48\textwidth]{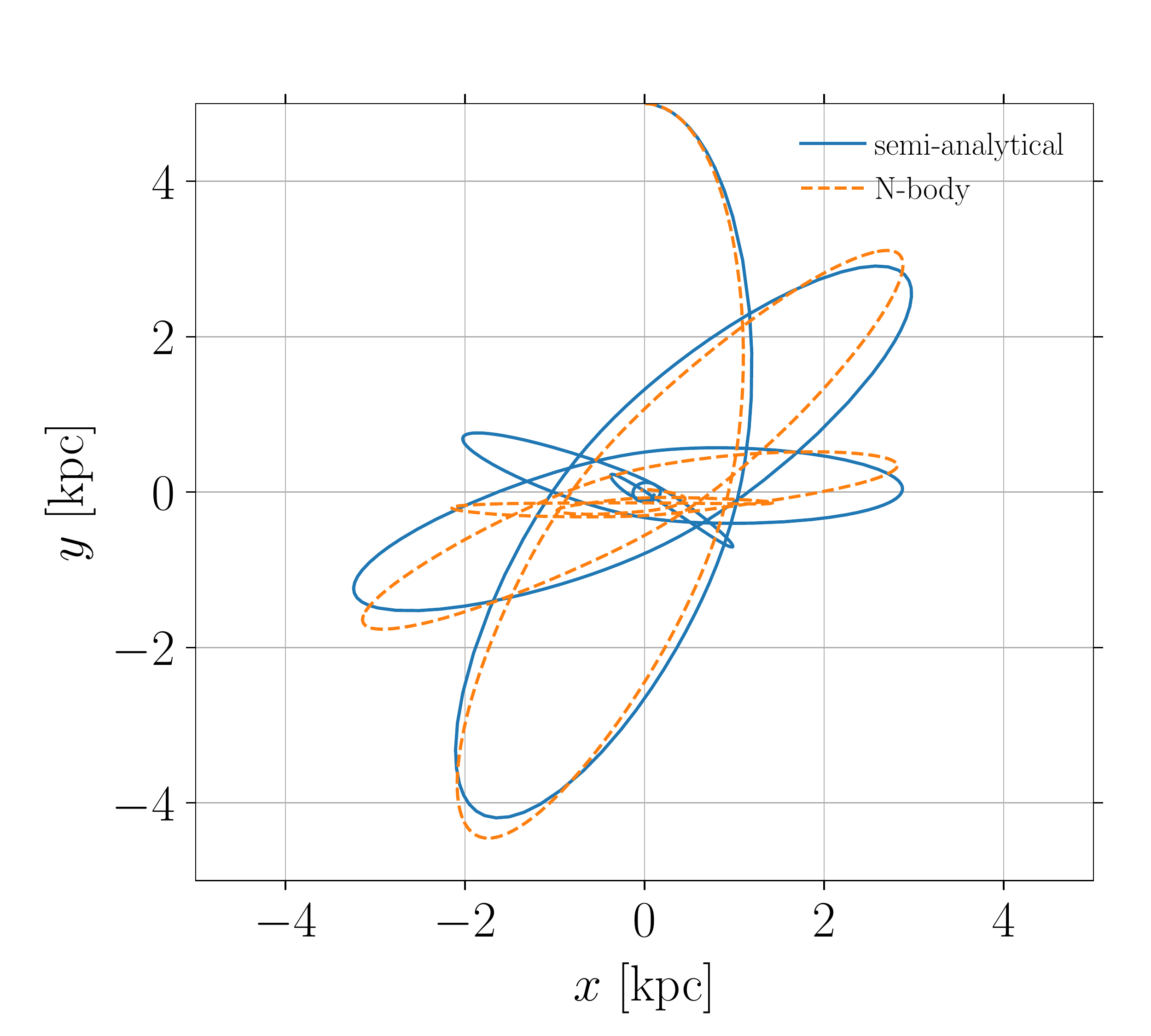}
    \caption{Same as Fig.~\ref{fig:prograde} but for an initially retrograde perturber.}
    \label{fig:retrograde}
\end{figure*}
%%%%%%%%%%%%%%%%%%

The choice of a massive bulge component was made mainly to enforce the Toomre-stability criterion over the whole stellar disc (as can be seen from the Toomre $Q$ parameter profile in Fig.~\ref{fig:ToomreQ}), thus suppressing the formation of strong local over-densities. 
Such features would add a stochastic forcing onto the perturber any time a close encounter takes place, hence contaminating  the net DTCC effect always present whenever the background density is rotating \citep[see][]{Fiacconi13,  delValle15, Roskar15, Lupi16, Tamburello17, SouzaLima17}. Our choice minimizes such stochastic perturbations and therefore allows for a clean test of the hypothesis that the DTCC effect is only due to DF.  

We create the initial conditions for the N-body runs using \textsc{GalIC} \citep{yurin14}, an iterative tool to create galaxy models in collisionless equilibrium, in which the particle orbits are integrated at each iteration for about ten orbital times.

We then set a MP with mass $M_{\rm MP} = 10^8\rm\,\msun$ at a separation $r_{\rm MP}=5\rm\, kpc$ from the galaxy centre, moving along an elliptic orbit in the disc equatorial plane with eccentricity $e_{\rm MP} = 0.7$. We consider two orbital configurations, with the MP either co-rotating or counter-rotating relative to the stellar disc. The mass and spatial resolution of the N-body simulations are chosen to guarantee that the dynamical friction onto the MP is properly resolved \citep[see the discussion in][]{Pfister2017}, as shown in Tab.~\ref{tab:paramters}.

In order to tune the free parameter $A_{\rm disc}$, we performed several tests to minimize the differences between the semi-analytical integration and the full N-body simulations, finding that the best value for $A_{\rm disc}$ is $\approx 0.5$. This best fit value, of the order of unity, suggests that our modeling catches all the major physical properties of disc DF. Hence, the final form of the disc DF acceleration used to obtain the results presented in this section is:
%%%%%%%%%%%%%%%%%%
\begin{equation}\label{eq:DF_final}
    \mathbf{a}_{\rm df,disc} = -\pi G^2 \ln(1+\Lambda^2) m_p \rho_d(R,0) \dfrac{\mathbf{v}_p-\mathbf{v}_{\rm circ}(R)}{|\mathbf{v}_p-\mathbf{v}_{\rm circ}(R)|^3},
\end{equation}
%%%%%%%%%%%%%%%%%%
with $\Lambda = p_{\rm max, d}/p_{\rm min, d}$.\footnote{During the perturber evolution we found that the average value of the factor containing the Coulomb logarithm (i.e. $A_{\rm disc} 2\pi \ln(1+\Lambda^2)$) lies around 20-22 for the prograde case and around 28-32 for the retrograde case. The difference can be ascribed to the higher relative velocity attained in the retrograde case. Additionally, in order to test the robustness of the employed prescription, we also tried to repeat the fitting procedure for $A_{\rm disc}$ with a slightly different galaxy with $z_d/R_d=0.15$ rather than 0.2 (see Tab.~\ref{tab:paramters}). Again we found that the best value for $A_{\rm disc}$ is $\approx 0.5$. Moreover, we also confirmed that $\Lambda$ remains practically unchanged with respect to the $z_d/R_d=0.2$ disc.}

Fig.~\ref{fig:prograde} and Fig.~\ref{fig:retrograde} compare the orbital evolution obtained via the full N-body runs (dashed orange lines) with
our semi-analytical prescription (solid blue lines). In both figures, the left-hand panels show the perturber radial separation\footnote{The high resolution of the N-body run ensures that the perturber orbit remains close to the disc equator.} (upper panel), its angular momentum normalised to its starting value (middle panel) and its orbital eccentricity\footnote{As long as the perturber lies in the equatorial plane, the potential depends on the radial coordinate only. This allows us to employ well known results valid for central potentials. At each time step we evaluate the total energy $E$ and angular momentum $h$ (per unit mass), and assuming them as instantaneously conserved, we obtain an expression for the radial velocity, i.e.
%%%%%%%%%%%%%%%%%%
\begin{equation}
    v_r^2 = 2\left(\dfrac{E}{m} - \phi(r)\right) - \dfrac{h^2}{r^2}.
\end{equation}
%%%%%%%%%%%%%%%%%%
Setting $v_r=0$, we can then numerically solve for $r$ to obtain the two roots that characterise bound orbits, i.e. the values of for the pericentre and apocentre, from which the eccentricity is ultimately evaluated.} (lower panel) as a function of time. The single right-hand panel instead reports the perturber motion in the galactic equatorial plane.

Fig.~\ref{fig:prograde} shows the orbital evolution for a prograde perturber, while in Fig.~\ref{fig:retrograde} we consider an initially retrograde case.
In spite of our simplistic assumptions, the semi-analytical evolution closely matches that of the full N-body simulation. This is particularly evident for the perturber radial separation, orbital eccentricity and motion in the disc plane. The angular momentum evolution is initially well reproduced, but, especially for the prograde case, it starts to deviate at later times. This behaviour is due to the assumption of locality made in our framework, which allows us to safely employ the semi-analytical prescription as long as the MP orbit is eccentric or, if circular, counter-rotating, while it becomes unreliable when the perturber orbit approaches a co-rotating circular one, i.e. when the relative velocity between the perturber and the disc at the perturber location goes to zero and equation~(\ref{eq:DF_final}) diverges. This limitation of our semi-analytical prescription is the reason why we limit ourselves to the study of the DTCC, and we do not focus on reproducing the whole evolution down to the centre of the composite system. Nevertheless, we can note that the orbital evolution is very well reproduced up to the point when the eccentricity approaches zero. Remarkably, as clear from the bottom left panels of Fig.~\ref{fig:prograde} and Fig.~\ref{fig:retrograde}, both for the prograde and retrograde cases, the time for which $e \rightarrow 0$ is practically the same of the corresponding N-body simulation.

Circularisation of initially retrograde orbits takes longer (by about a factor of two) with respect to the prograde case (as shown by the comparison between Fig.~\ref{fig:prograde} and  Fig.~\ref{fig:retrograde}). The difference in the timescales is due to the fact that, in order to circularise the MP orbit, DF has first to reverse the MP angular momentum, forcing the retrograde configuration into a prograde one. Still, despite the simplistic assumptions made, the time at which the sign reversal happens turns out to be quite similar to that of the N-body run and even more remarkably the eccentricity evolution gets closely reproduced up to the time of circularisation. We stress again that our goal is to model the motion of eccentric perturbers (or circular but counter-rotating), where velocities relative to the stellar background are non-negligible. In this regime our approach closely follows the outcomes of N-body simulations confirming that the angular momentum reversal, as the circularisation, is determined by DF.  

%%%%%%%%%%%%%%%%%%%%%%%%%%%%%%%%%%%
\section{Discussion and conclusions}
\label{sec:discussion}

In this work we considered a simple semi-analytical approach to describe the motion of MPs inside rotating axisymmetric structures. We validated our prescription against full N-body simulations, finding a remarkable good agreement.
We verified that DF drives the MP motion towards circular orbits, even when initially the MP counter-rotates. In this last case, first the MP angular momentum gets reversed causing the MP to co-rotate with the disc, then, once on a prograde orbit, DF circularises it. 

Here, we focused on rotating flat structures, but this feature does not specifically require a disc-like geometry. The DTCC effect is indeed common in any rotating structure and it could play an influential role even in fast rotating ellipticals \citep{Emsellem2011} or galactic nuclei \citep[see e.g. the discussion in][]{Sesana11,Rasskazov17}. Still, since a proper analytical description for such cases is not as easy as for the current case, we do not explore that further, limiting our discussion to the disc case only.

In addition to the consequences of the DTCC effect for massive black hole binaries (discussed in the Introduction and in the references therein), the DTCC also has a direct impact on the properties of kinematically decoupled stellar cores (KDCs). KDCs are central stellar components with a rotation axis that is misaligned from the main stellar body of the galaxy. They are characterised by an abrupt change of more than 30 degrees in the kinematic position angle and are observed in a significant fraction of early-type galaxies ($\sim7\%$ \citealt{krajnovic11}). Their presence is a clear signature of an external accretion event, with smaller KDCs (diameter $<$ 1 kpc) associated with more recent accretion events, either caused by gas accretion followed by subsequent in-situ star formation or caused by a minor merger \citep{mcdermid06, raimundo13}.

In general, one could expect that if a minor merger occurs with a random orientation, the probability of observing decoupled cores that co-rotate or counter-rotate with respect to the main stellar body of the galaxy would be approximately similar. However, this does not seem to be the case, at least for the distribution of externally accreted gas. \citet{davis11} find a larger fraction of co-rotating than counter-rotating gas in early type galaxies, specifically in a $\sim$ 65/35 ratio between co-rotating gas and gas with large misalignment ($>$ 30 deg). It is still not clear why such trend is observed, but some suggested reasons are that minor merger accretion may be anisotropic, occurring along a preferential direction such as the major axes of the dark matter halo \citep{deason11, davis11}, or that there is a drag towards co-rotation due to the interaction with already present gas \citep[e.g. in the hot halo of the galaxy,][]{davis11}.
If a KDC forms from the gas that has been externally accreted, then the statistics for the stellar core misalignment should be the same as for the gas since they would share similar dynamics.

The work presented here suggests an alternative mechanism to promote the higher relative occurrence of co-rotating KDSc. Our work predicts that for galaxies with discs, such as S0s and spiral galaxies, DF can result in the inversion of the angular momentum of a massive perturber, from a counter-rotating orbit to a co-rotating orbit. This is especially relevant for minor mergers, where we expect the accreted satellite galaxy to be a massive perturber and at a later stage to form a KDC. If one considers that the original direction of the minor merger is randomly oriented, this angular momentum inversion would result in a larger fraction of co-rotating KDCs than counter-rotating KDCs formed after minor merger events. 
The consequence of this alternative mechanism is an overall tendency for co-rotating KDCs in galaxies with discs, and its relevance can be observationally tested with the study of larger samples of KDCs and with the characterization of the host IGM properties.
The fact that the DF mechanism studied here operates even in extremely gas poor systems, makes it a an important mechanism to consider even in early-type galaxies with low native gas content. In addition, after circularisation, DF removes efficiently angular momentum leading the perturber to the host nucleus without requiring a finely tuned initial angular momentum to justify the fall of externally accreted material down to the inner kpc of its host.

In analogy with KDCs, DTCC could have a similar impact on the kinematic properties of nuclear star clusters. Such systems have been proposed to form either via in-situ star formation \citep[e.g.][]{Loose82}, and/or via the accretion of multiple stellar clusters reaching the centre of the system via DF \citep[e.g.][]{Tremaine75}.
The in-situ channel has been always regarded as more promising if one wants to explain the kinematic properties of nuclear star clusters, and in particular their significant rotation. On the other hand, recently, \citet{Tsatsi17} showed that even the formation via clusters accretion can bring to a net significant rotation, even if the orbits of the inspiralling clusters are isotropic. Our result strengthens their findings as, if we account for DTCC, clusters coming from the galactic disc should be accreted with the same direction in angular momentum (if their circularisation occurs before the cluster gets tidally disrupted).

Finally, in a future study we plan to expand our semi-analytical treatment to the case of circular perturbers (corresponding to the final configuration of any initial configuration as a consequence of the DTCC effect) and to out-of-plane geometries. This prescription will be instrumental in order to fully characterize the DF time distribution of infalling perturbers as a function of their host galaxy properties.

%%%%%%%%%%%%%%%%%%%%%%%%%%%%%%%%%%%%%%%%%%%%%%%%%%%%%%%%%%%%%%%%%%%%%%%%%%%%%%%%%%%%%
%%%%%%%%%%%%%%%%%%%%%%%%%%%%%%%%%%%%%%%%%%%%%%%%%%%%%%%%%%%%%%%%%%%%%%%%%%%%%%%%%%%%%
\section*{Acknowledgements}

Numerical calculations have been made possible through a CINECA-INFN agreement, providing access to resources on GALILEO and MARCONI at CINECA.
AL acknowledges support from the European Research Council No. 740120 `INTERSTELLAR'. 
EB acknowledges support from  the \textit{Swiss National Science Foundation} under the Grant 200020\_178949. 
S.I.R. gratefully acknowledges support from the Independent Research Fund Denmark via grant numbers DFF 4002-00275 and 8021-00130. 
MD acknowledges Giuseppe Lodato for pointing out that the expression "angular momentum flip" could be misleading (see footnote~\ref{footnote:1}). 

%%%%%%%%%%%%%%%%%%%%%%%%%%%%%%%%%%%%%%%%%%%%%%%%%%
%%%%%%%%%%%%%%%%%%%% REFERENCES %%%%%%%%%%%%%%%%%%

% The best way to enter references is to use BibTeX:

\bibliographystyle{mnras}
\bibliography{biblio} % if your bibtex file is called biblio.bib

%%%%%%%%%%%%%%%%%%%%%%%%%%%%%%%%%%%%%%%%%%%%%%%%%%
%%%%%%%%%%%%%%%%% APPENDICES %%%%%%%%%%%%%%%%%%%%%

% \clearpage
% \appendix
% \section{Appendix}

%%%%%%%%%%%%%%%%%%%%%%%%%%%%%%%%%%%%%%%%%%%%%%%%%%

% Don't change these lines
\bsp	% typesetting comment
\label{lastpage}
\end{document}